\documentclass{ws-ijmpcs}

\newcommand{\abs}[1]{\vert #1\vert}
\newcommand{\Det}[1]{\left\vert #1\right\vert}
\begin{document}

\markboth{Pablo Rodriguez-Lopez}
{Casimir Energy and Entropy between perfect metal Spheres}

%
\catchline{}{}{}{}{}
%

\title{Casimir Energy and Entropy between perfect metal Spheres}

\author{PABLO RODRIGUEZ-LOPEZ}

\address{Departamento de F\'\i sica Aplicada I and GISC,\\
Facultad de Ciencias F\'\i sicas, Universidad Complutense, 28040 Madrid, Spain and
Departamento de Matem\'aticas and GISC, Universidad Carlos III de Madrid,
Avenida de la Universidad 30, 28911 Legan\'es, Spain\\
pablo.rodriguez@fis.ucm.es}

\maketitle

\begin{history}
\received{Day Month Year}
\revised{Day Month Year}
\end{history}

\begin{abstract}
We calculate the Casimir energy and entropy for two perfect metal spheres in the large and short separation limit. We obtain nonmonotonic behavior of the Helmholtz free energy with separation and temperature, leading to parameter ranges with negative entropy, and also nonmonotonic behavior of the entropy with temperature and with the separation between the spheres. The appearance of this anomalous behavior of the entropy is discussed as well as its thermodynamic consequences.

\end{abstract}

\ccode{PACS numbers: 03.70.+k, 12.20.−m, 42.50.Lc, 85.85.+j}


\section{Introduction}
In 1948, Casimir predicted the attraction between perfect metal parallel plates\cite{Casimir Placas Paralelas} due to quantum fluctuations of the electromagnetic field. Recently, a multiscattering formalism of the Casimir effect for the electromagnetic field has been presented\cite{EGJK,Review-Jamal-Emig,Multiscattering_Lambrecht,Multiscattering_Milton}.

The Casimir effect has some peculiarities. In particular, it is a non-pairwise interaction; the Casimir thermal force (the thermal part of the Casimir energy) between two isolating bodies is not necessarily monotonic in their separation, as seen in the sphere--plate and cylinder--plate cases\cite{Metodo_Caminos_Esfera_Placa_Escalar}. In addition, for some geometries, intervals of negative entropy appear, as in the case of two parallel plates described by the Drude model\cite{Entropias_negativas_placas_Drude} or, as recently shown, in the interaction between a Drude model plate and sphere\cite{Bordag_Entropia_Esfera_Placa} and in the interaction between a perfect metal plate and sphere\cite{Canaguier-Durand_Caso_Esfera_Placa_PRA}.

In this article we study the Casimir effect between two equal radii perfect metal spheres in the large and short distance limit. As a result, we find negative entropies in certain ranges of temperature and separation between the spheres. In addition, we find nonmonotonic behavior of the entropy with the separation while the force is attractive for all separations, making it appear as though the natural evolution of the system tends to increase the entropy in certain ranges of temperature and separation. We discuss the thermodynamical meaning and consequences of negative entropies in Casimir effect. Similar results for spheres with more general dielectric models can be found in Ref.~\refcite{PRB_Negative_Entropy_2_Spheres} and for Drude plates in Ref.~\refcite{Brevik} .

The remainder of the article is arranged as follows: In Sect. \ref{Presentacion del modelo de Emig del Efecto Casimir}, we describe the multiscattering model used herein to obtain the Casimir energies and entropies for the two perfect metal spheres. In Sect. \ref{Casimir energy}, we obtain the large and short separation limit of the Casimir energy,
In Sect. \ref{Casimir entropy} we obtain the entropy of the system, analytically and numerically. In Sect. \ref{Casimir force} we obtain the Casimir force.
We discuss the thermodynamic consequences of these results in Sect. \ref{Sect:Thermodynamic}. Finally, we discuss the results obtained in the Conclusions.

\section{Multiscattering formalism of the Casimir energy}\label{Presentacion del modelo de Emig del Efecto Casimir}
To calculate the Casimir energy, entropy, and forces between two spheres, we employ the multiscattering formalism for the electromagnetic field\cite{EGJK,Review-Jamal-Emig}. The Casimir contribution to the Helmholtz free energy at any temperature $T$ is given by
\begin{equation}\label{Energy_T_finite}
E = k_{B}T{\sum_{n=0}^{\infty}}'\log\Det{\mathbb{I} - \mathbb{N}(\kappa_{n})},
\end{equation}
where $\kappa_{n} = \frac{n}{\lambda_{T}}$ are the Matsubara frequencies and $\lambda_{T} = \frac{\hbar c}{2\pi k_{B}T}$ is the thermal wavelength. The prime indicates that the zero Matsubara frequency contribution has height of $1/2$. All the information regarding the system is described by the $\mathbb{N}$ matrix. For a system of two objects, this matrix is $\mathbb{N} = \mathbb{T}_{1}\mathbb{U}_{12}\mathbb{T}_{2}\mathbb{U}_{21}$. $\mathbb{T}_{i}$ is the T scattering matrix of the $i^{\text{\underline{th}}}$ object, which accounts for all the geometrical information and electromagnetic properties of the object. $\mathbb{U}_{ij}$ is the translation matrix of electromagnetic waves from object $i$ to object $j$, which accounts for all information regarding the relative positions between the objects of the system.

For a perfect metal sphere of radius $R$, the $\mathbb{T}$ matrix is diagonal in $\left(l m P,l' m'P'\right)$ space\cite{Review-Jamal-Emig}, with elements given by
\begin{equation}
\mathbb{T}^{MM}_{l m,l' m'} = - \delta_{l\,l'}\delta_{mm'}\frac{\pi}{2}\frac{I_{l + \frac{1}{2}}(\kappa R)}{K_{l + \frac{1}{2}}(\kappa R)},
\end{equation}
\begin{equation}
\mathbb{T}^{EE}_{l m,l' m'} \hspace{-2pt}=\hspace{-2pt} - \delta_{l\,l'}\delta_{mm'}\frac{\pi}{2}\frac{l I_{l + \frac{1}{2}}(\kappa R) - \kappa R I_{l - \frac{1}{2}}(\kappa R)}{l K_{l + \frac{1}{2}}(\kappa R) + \kappa R K_{l - \frac{1}{2}}(\kappa R)}.
\end{equation}
Expressions for the $\mathbb{U}_{\alpha\beta}$ matrices can be found in Refs.~\refcite{Review-Jamal-Emig} and \refcite{Wittmann}.

\section{Casimir energy}\label{Casimir energy}
\subsection{Large separation limit}
\begin{figure}
\begin{center}
\includegraphics[scale=0.65]{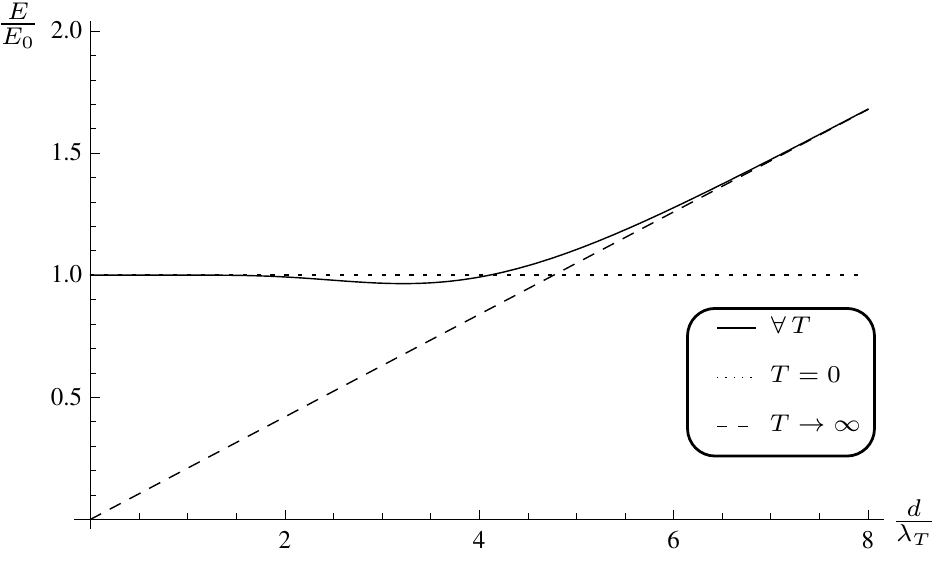}
\includegraphics[scale=0.65]{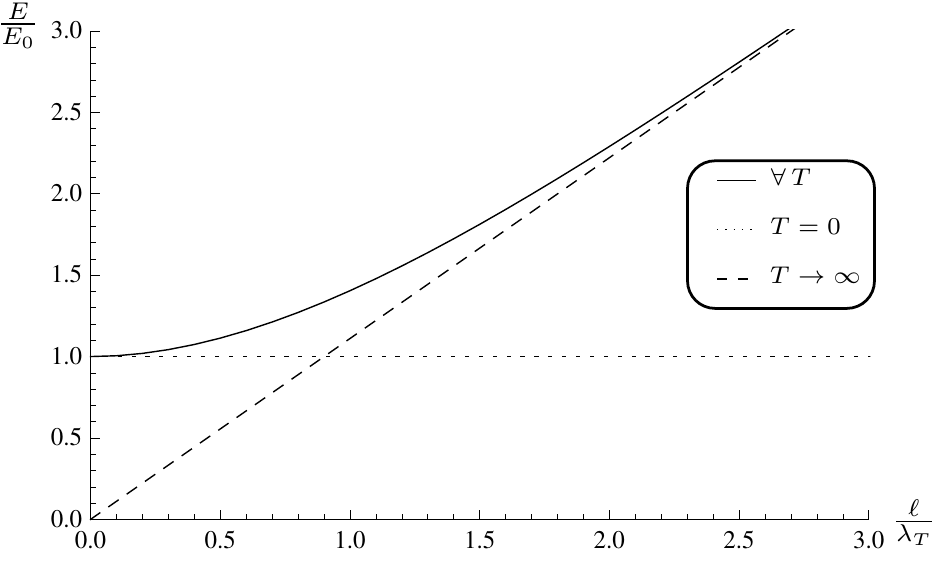}
\caption{The left and right figures are the large ($d,\ell \gg R$) and short ($\ell \ll R$) distance limit of the Casimir energy between perfect metal spheres as a function of $\frac{d}{\lambda_{T}}$ (left) and of $\frac{\ell}{\lambda_{T}}$ (right) compared with the energy at zero temperature respectively. The dotted curve is the quantum limit, the dashed curve is the classical limit, and the solid curve is the asymptotic finite-temperature Casimir energy. Note that these curves are independent of the radius of the spheres.}
\label{Energia_2_esferas_metalicas}
\end{center}
\end{figure}

To obtain the large separation limit of the Casimir energy, we need the dominant part of the $\mathbb{T}$ matrix in this limit. We define the adimensional frequency $q$ by $\kappa = q/d$, where $d$ is the distance between the centre of the spheres. The main contribution in the large separation limit comes from the lowest-order expansion term of the $\mathbb{T}$ matrix elements in $1/d$. The dominant contribution comes from the dipolar polarizabilities part of the $\mathbb{T}$ matrix, taking the form
\begin{equation}
\mathbb{T}^{MM}_{1 m,1 m'} = - \frac{1}{3}\left(\frac{qR}{d}\right)^{3},
\hspace{50pt}
\mathbb{T}^{EE}_{1 m,1 m'} =   \frac{2}{3}\left(\frac{qR}{d}\right)^{3}.
\end{equation}
Using the property $\log\abs{A} = \text{Tr}\log(A)$, and expanding Eq. \eqref{Energy_T_finite} over $\frac{1}{d}$, we obtain
\begin{equation}
E = - k_{B}T{\sum_{n = 0}^{\infty}}'\text{Tr}\left(\mathbb{N}(\lambda_{T}^{-1}n)\right).
\end{equation}
Using the translation matrices in a spherical vector multipole basis\cite{Review-Jamal-Emig} and the large separation approximation of the $\mathbb{T}$ matrix, the trace of the $\mathbb{N}$ matrix is obtained by straightforward calculus.
Here we denote with a sub--index $T$ the results valid for all temperatures, with a sub--index $0$ the results in the quantum limit ($T\to 0$), and with a sub--index $cl$ the results in the classical limit ($\hbar\to 0$), which is equivalent to the high $T$ limit.
Carrying out the sum over Matsubara frequency, we obtain the Casimir contribution to the Helmholtz free energy for two spheres of equal radius $R$ as
\begin{eqnarray}\label{Energia_finite_T_metal_perfecto}
E_{T} & = & - \frac{\hbar c R^{6}}{2\pi d^{7}}\frac{z e^{5z}}{2\left(e^{2z} - 1\right)^{5}}\times\nonumber\\
& & \left(2\left(15 - 29z^{2} + 99z^{4}\right)\cosh(z) + 15\cosh(5z) + \left( - 45 + 58z^{2} + 18z^{4}\right)\cosh(3z)\right.\nonumber\\
& & \left. + 24z\left( 6z^{2} - 5 + \left(5 + 3z^{2}\right)\cosh(2z)\right)\sinh(z)\right),
\end{eqnarray}
where $z = d/\lambda_{T}$. We also define the adimensional Casimir energy as $E_{ad}(z) = \frac{2\pi d^{7}}{\hbar c R^{6}}E_{T}$. From this result, the quantum ($T\to 0$) and classical ($\hbar\to 0$) limits with their first corrections are easily obtained as
\begin{equation}\label{Low_T_corrections_Energy}
E_{0} = - \frac{143 \hbar c R^{6}}{16\pi d^{7} } - \frac{8 \hbar c\pi^{5}R^{6}}{27 d}\left(\frac{k_{B}T}{\hbar c}\right)^{6} + \frac{2288 d \hbar c\pi^{7}R^{6}}{1575}\left(\frac{k_{B}T}{\hbar c}\right)^{8},
\end{equation}
\begin{eqnarray}\label{High_T_corrections_Energy}
E_{cl} & = & - \frac{15 k_{B}T R^{6}}{4 d^{6}} - k_{B}T\frac{R^{6}}{2d^{6}}\left(15 + 30 z + 29 z^{2} + 18 z^{3} + 9 z^{4}\right)e^{-2z}.
\end{eqnarray}
Note that in Eq. \eqref{Low_T_corrections_Energy}, the first correction to zero temperature case is proportional to $T^{6}$, contrary to the plate--sphere case\cite{Metodo_Caminos_Esfera_Placa_Escalar,Canaguier-Durand_Caso_Esfera_Placa_PRA} and to the cylinder--plate case\cite{Metodo_Caminos_Esfera_Placa_Escalar}, where a result proportional to $T^{4}$ were obtained for both systems. As corrections have the same negative sign of the Casimir energy in both limits, they describe an increase of the magnitude of the Casimir energy in high and low temperature limits. It is no longer the case if we study the energy for all temperatures. In fact, for some ranges of separation and temperature, thermal photons tend to reduce the Casimir energy between the spheres, as shown in the left plot of  Fig. \ref{Energia_2_esferas_metalicas}, which is an indicator of the appearance of negative entropy in this system because of the negative slope of the energy curve.
It is clear in Fig. \ref{Energia_2_esferas_metalicas}(left) that for some temperatures and distances, the Casimir energy between spheres is lower than in the zero temperature case. But it is less evident the validity of Eq. \eqref{Low_T_corrections_Energy}. In fact, there is a tiny increase of $E/E_{0}$ at low $\frac{d}{\lambda_{T}}$ until reaching a local maximun at $\frac{d}{\lambda_{T}}\approx 1.0388$ of $E/E_{0}\approx 1 + 10^{-4}$. It is not visible in Fig. \ref{Energia_2_esferas_metalicas} because of the difference of scales. 

\subsection{Short distance limit, PFA}
To study the short distance limit of our problem, we apply the Proximity Force Approximation (PFA)\cite{Teo_PFA,Post_PFA_original,Post_PFA_Emig}. We start with the general formula of Casimir energy between parallel plates for all temperature $T$ as
\begin{equation}\label{PFA_2}
E^{T}_{\parallel} = - \frac{k_{B}T A}{2\pi}{\sum_{n=0}^{\infty}}'\sum_{m=1}^{\infty}\frac{1}{m}\int_{0}^{\infty}dk_{\perp}\,k_{\perp}\text{Tr}\left(R_{1}\cdot R_{2}\right)e^{-2md\sqrt{ k_{\perp}^{2} + \lambda_{T}^{-2}n^{2} }}.
\end{equation}
PFA is a valid approximation at short distances. Following Ref.~\refcite{Teo_PFA}, we define the distance between equal radius $R$ spheres as $h(\theta, \varphi) = \ell + 2R(1 - \cos(\theta))$, where $\ell$ is the minimum distance between the surfaces of the spheres and $\rho = R\sin(\theta)$, then we calculate the PFA energy as the short distance limit of the next integral
\begin{equation}\label{PFA_3}
E^{PFA}_{\circ\circ} = \int_{0}^{2\pi}d\varphi\int_{0}^{R}d\rho\,e^{T}_{\parallel}(\ell + 2R(1 - \cos(\theta)),
\end{equation}
where $e_{\parallel} = \frac{dE_{\parallel}}{dA}$. Carrying out the integration over $\varphi$ and applying the change $t = 1 - \cos(\theta)$,
\begin{equation}\label{PFA_4}
E^{PFA}_{\circ\circ} = 2\pi R^{2}\int_{0}^{1}dt( t - 1 ) e^{T}_{\parallel}(\ell + 2Rt).
\end{equation}
To separate the divergence of the PFA energy, we apply the change of variable $t = \frac{\ell}{R}u$ and take the short distance limit ($\ell\to 0$), leading to our final PFA result,
\begin{equation}\label{PFA_6}
E^{PFA}_{T} = - 2\pi R\ell\int_{0}^{\infty}du e^{T}_{\parallel}(\ell( 1 + 2u)).
\end{equation}
The formula presented above is a valid approximation for all temperatures and material properties of the equal radius spheres at a short distance $\ell$. For perfect metal spheres we have $\text{Tr}\left(R_{1}\cdot R_{2}\right) = 2$, then the $k_{\perp}$ integral of Eq.~\eqref{PFA_2} can be carried out and the obtained result is applied to Eq.~\eqref{PFA_6}, obtaining
\begin{equation}\label{PFA_8}
E^{PFA}_{T} = - \frac{k_{B}TR}{4\ell}\sum_{m=1}^{\infty}{\sum_{n=0}^{\infty}}'\frac{e^{- 2nm\ell\lambda_{T}^{-1}}}{m^{3}}.
\end{equation}
The quantum\cite{EGJK} and classical limits are
\begin{equation}\label{PFA_9}
E^{PFA}_{0} = - \frac{\hbar c\pi^{3}R}{1440\ell^{2}} - \frac{k_{B}T\pi^{3}R}{72}\left(\frac{k_{B}T}{\hbar c}\right) + \frac{k_{B}T\ell R\zeta(3)}{2}\left(\frac{k_{B}T}{\hbar c}\right)^{2} + \cdots
\end{equation}
\begin{equation}\label{PFA_10}
E^{PFA}_{cl} = - \frac{k_{B}TR\zeta(3)}{8\ell} - \frac{k_{B}TR}{4\ell}e^{- 4\pi\ell \frac{k_{B}T}{\hbar c}} + \cdots
\end{equation}
In Fig. \ref{Entropia_2_esferas_metalicas}(right), the energy in the short distance limit is shown as a function of $\frac{\ell}{\lambda_{T}}$. In this limit the energy grows monotonically with T, so we do not expect to find negative entropies in this limit.

\section{Casimir entropy}\label{Casimir entropy}
\begin{figure}
\begin{center}
\includegraphics[scale=0.68]{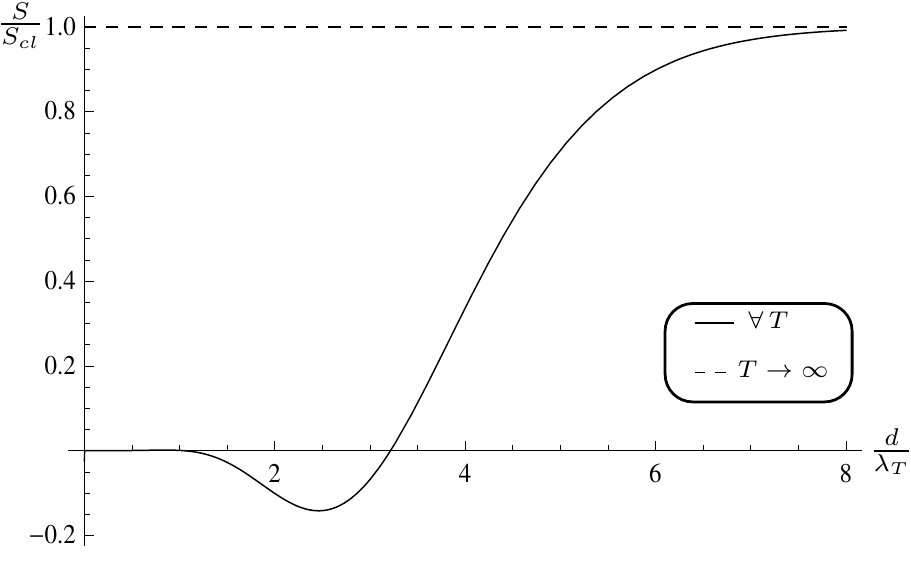}
\includegraphics[scale=0.68]{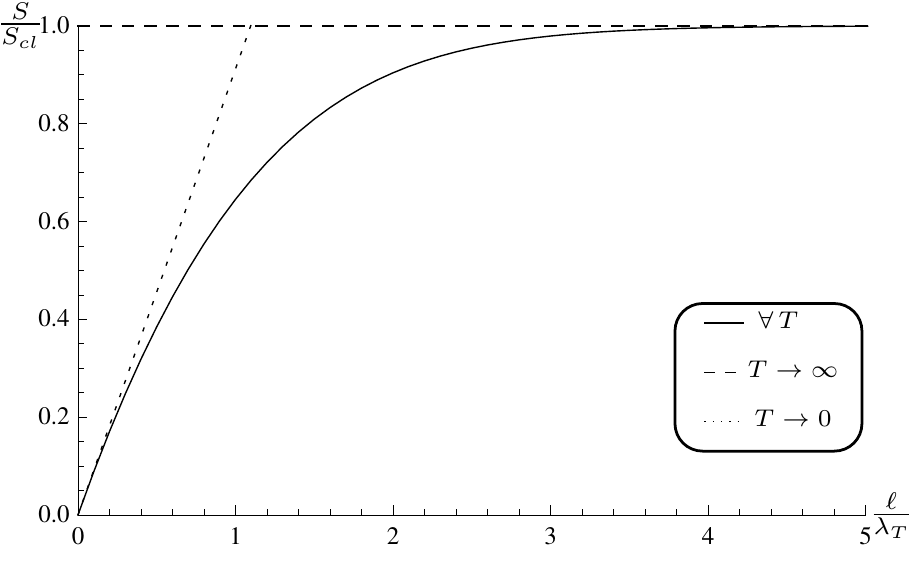}
\caption{The left and right figures are the large and short distance limit of the Casimir entropy between perfect metal spheres as a function of $\frac{d}{\lambda_{T}}$ (left) and of $\frac{\ell}{\lambda_{T}}$ (right) compared with the classical limit respectively. The dotted curve is the first temperature correction of the quantum limit entropy (the linear term of Eq.~\eqref{PFA_12}), the dashed curve is the classical limit, and the solid curve is the asymptotic finite-temperature Casimir entropy. Note the regions of negative entropy and negative slope of the entropy with the parameter $\frac{d}{\lambda_{T}}$ in the large distance limit dissappears in the short distance case.}
\label{Entropia_2_esferas_metalicas}
\end{center}
\end{figure}

In the canonical ensemble, the entropy is defined as $S = - \partial_{T}E$. In the large separation limit, the Helmholtz free energy depends on the adimensional variable $z = \frac{d}{\lambda_{T}} = 2\pi\frac{dk_{B}T}{\hbar c}$, so we can write the entropy as
\begin{equation}
S = - \frac{\partial z}{\partial T}\frac{\partial E}{\partial z} = - 2\pi\frac{dk_{B}}{\hbar c}\frac{\partial E}{\partial z},
\end{equation}
and define the adimensional entropy as $S_{ad}(z) = \frac{d^{6}}{k_{B}R^{6}}S = - \partial_{z}E_{ad}(z)$. From this result, the quantum ($T\to 0$) and classical ($\hbar\to 0$) limits are easily obtained as
\begin{equation}\label{Low_T_Entropy}
S_{0} = 0 + \frac{16  k_{B} \pi^{5}R^{6} }{9 d }\left(\frac{k_{B}T}{\hbar c}\right)^{5} - \frac{18304 k_{B}d \pi^{7}R^{6}}{1575}\left(\frac{k_{B}T}{\hbar c}\right)^{7},
\end{equation}
\begin{eqnarray}\label{High_T_Entropy}
S_{cl} & = & \frac{15 k_{B} R^6}{4 d^6} + k_{B}\frac{R^{6}}{2d^{6}}\hspace{-1pt}\left(15 \hspace{-1pt}+\hspace{-1pt} 30z \hspace{-1pt}+\hspace{-1pt} 27z^{2} \hspace{-1pt}+\hspace{-1pt} 14z^{3} \hspace{-1pt}+\hspace{-1pt} 9z^{4} \hspace{-1pt}-\hspace{-1pt} 18z^{5}\right)\hspace{-1pt}e^{-2z}\hspace{-2pt},
\end{eqnarray}
where $z = \frac{d}{\lambda_{T}}$. So, the entropy is a growing function with temperature in both limits, but this is not the case for all temperatures, as we can observe in the left plot of Fig. \ref{Entropia_2_esferas_metalicas}, where a region of negative entropy and another region of negative slope of the entropy are observed.

Because of the limit at low temperature of the entropy, we know that the entropy is positive for low $T$ (not seen in Fig. \ref{Entropia_2_esferas_metalicas} because it is small compared with $S/S_{\text{cl}}$, but it can be observed in Fig. \ref{LogLogPlot_Entropia_2_esferas_metalicas_forall_distances}), so there are three points where $S=0$, including the origin. 

Negative entropy of the Casimir effect has already been obtained between Drude parallel plates in Ref.~\refcite{Entropias_negativas_placas_Drude} and in Ref.~\refcite{Ingold}, and more recently between a perfect metal plate and sphere in Ref.~\refcite{Canaguier-Durand_Caso_Esfera_Placa_PRA} and between a Drude sphere and plate in Ref.~\refcite{Zandi_Emig_Placa_Esfera_varios_modelos} and in Ref.~\refcite{Bordag_Entropia_Esfera_Placa}.

These results are only valid when the separation between the spheres is large compared with their radius, regardless of the radius of each one.

In the Short distance limit, the entropy is directly obtained from Eq.~\eqref{PFA_8} as
\begin{equation}\label{PFA_11}
S^{PFA}_{T} = \frac{k_{B}R}{4\ell}\sum_{m=1}^{\infty}{\sum_{n=0}^{\infty}}'\frac{e^{- 2nm\ell\lambda_{T}^{-1}}}{m^{3}}\left(1 - 2nm\ell\lambda_{T}^{-1}\right),
\end{equation}
and the high and low temperature limits are
\begin{equation}\label{PFA_12}
S^{PFA}_{0} = 0 + \frac{k_{B}\pi^{3}R}{36}\left(\frac{k_{B}T}{\hbar c}\right) - \frac{3k_{B}\ell R\zeta(3)}{2}\left(\frac{k_{B}T}{\hbar c}\right)^{2} + \cdots,
\end{equation}
\begin{equation}\label{PFA_13}
S^{PFA}_{cl} = \frac{k_{B}R\zeta(3)}{8\ell} - k_{B}R\pi\frac{k_{B}T}{\hbar c}e^{-4\pi\ell\frac{k_{B}T}{\hbar c}} + \cdots,
\end{equation}
then the entropy is positive for all temperatures in the short distance limit. We can observe in Fig. \ref{Entropia_2_esferas_metalicas}(right) that the negative entropy interval dissappears at short distances and, at low temperature, the entropy is linear with $T$, contrary to the far distance limit, where a slope of $T^{5}$ was obtained.

\subsection{Numerical study at smaller separations}
As noted previously, asymptotic results are no longer valid when the separation between the spheres becomes comparable to their radius. For this reason, a numerical study of entropy was performed for these cases. We computed Eq. \eqref{Energy_T_finite} numerically for all temperatures from $T=0$ until reaching the classical limit for fixed ratio between the radius $R$ and separation, $r = \frac{R}{d}$.



\begin{figure}
\begin{center}
\includegraphics[scale=0.74]{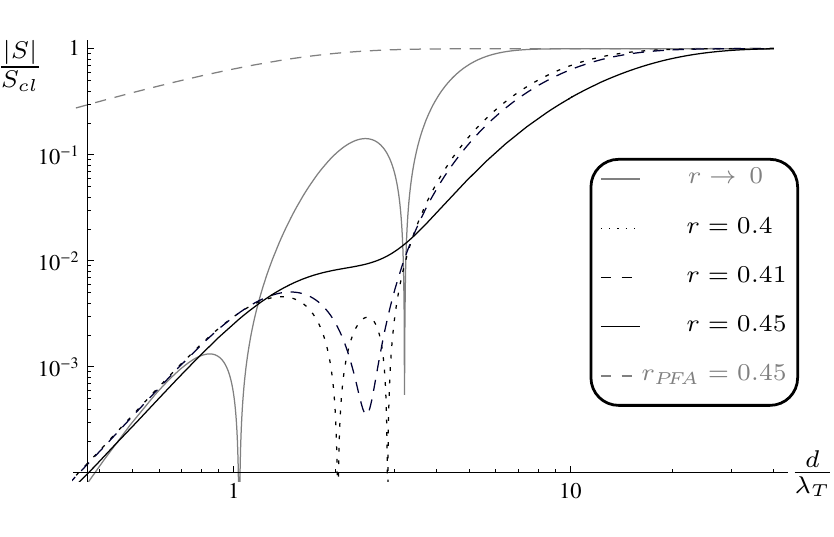}
\includegraphics[scale=0.74]{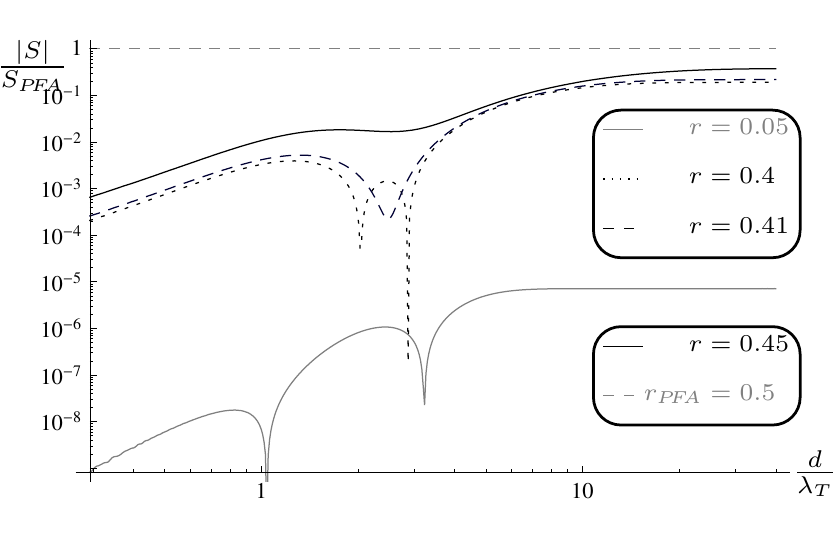}
\caption{The left figure is the log--log plot of the absolute value of the entropy divided by its classical limit for perfect metal spheres at constant $r$ as a function of $\frac{d}{\lambda_{T}}$. Starting from the asymptotic solid gray curve ($r\to 0$), we increase $r$ up to the dotted curve $r = 0.4$, observing a reduction of the interval of negative entropy. Dashed curve is the case $r = 0.41$, where the interval of negative entropy has disappeared. The solid black curve for $r=0.45$ is shown and the PFA result for $r=0.45$ is the dashed gray curve. The right figure is the log--log plot of the absolute value of the entropy divided by the PFA limit for perfect metal spheres at constant $r$ as a function of $\frac{d}{\lambda_{T}}$ too. In this case, the solid gray curve represents $r= 0.05$ instead the asymptotic far distance result.}
\label{LogLogPlot_Entropia_2_esferas_metalicas_forall_distances}
\end{center}
\end{figure}

In Fig. \ref{LogLogPlot_Entropia_2_esferas_metalicas_forall_distances}(left), the entropy of the system of two perfect metal spheres is plotted as a function of $z = \frac{d}{\lambda_{T}}$ for constant $r = \frac{R}{d}$. The large and short separation limit result are shown too. We choose a log--log representation of the absolute value of the entropy divided by its corresponding classical limit. Therefore, zeros are observed as log divergences, and we can also observe the cases of negative entropy. Starting in the large separation regime, we observe an interval of negative entropy. As we increase $r$ (reducing the separation between the spheres), the region of negative entropy tends to reduce until it disappears between $r = 0.40$ and $r = 0.41$. The power-law decay of the entropy at low temperatures  (Eq.~\eqref{Low_T_Entropy}) is observed as a linear decay of the curve at low $z$, and constant behavior in the high-temperature limit (Eq.~\eqref{High_T_Entropy}) is also reached in the computation.

In the right plot of Fig. \ref{LogLogPlot_Entropia_2_esferas_metalicas_forall_distances}, we observe the ratio of the entropy with the PFA limit also as a function of $z = \frac{d}{\lambda_{T}}$ for constant $r = \frac{R}{d}$. In this case we see that the numerical results tend to the PFA limit when spheres tend to contact on a non--trivial way, with a better convergence for high $z$ than for low $z$. The power law decay of the entropy at low temperatures reduces from $S\propto T^{5}$ for the large distance result ($r\to 0$), around $S\propto T^{3}$ for the closest numerical studied case ($r = 0.45$) to $S\propto T$ for the short distance limit. This polynomial behavior $S\propto T^{\alpha(r)}$ is observed in Fig. \ref{LogLogPlot_Entropia_2_esferas_metalicas_forall_distances} as a linear decay of the curves at low $z$. As a consequence, the convergence of the entropy to the PFA result at low temperature will be reached at extremely short distances between the spheres.

\section{Casimir force}\label{Casimir force}
\begin{figure}
\begin{center}
\includegraphics[scale=0.65]{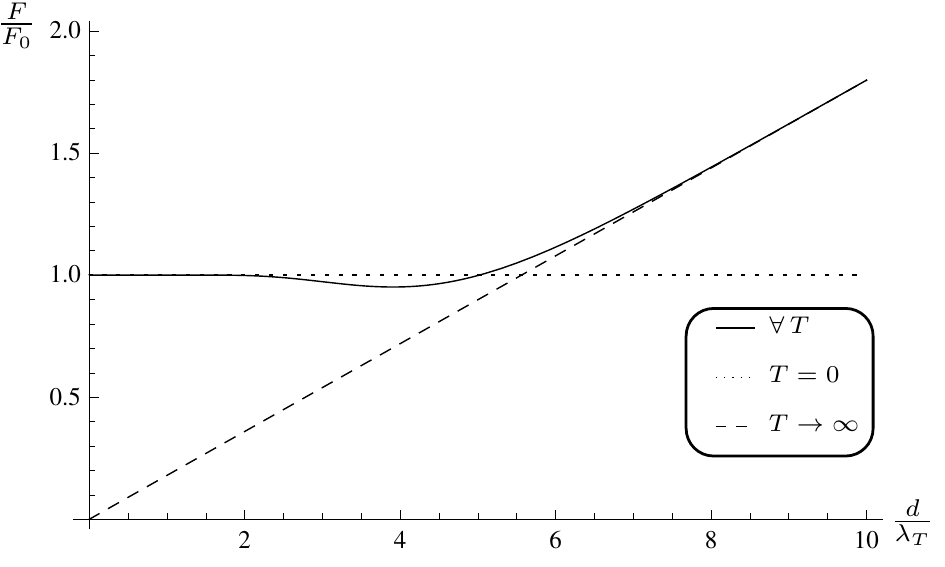}
\includegraphics[scale=0.65]{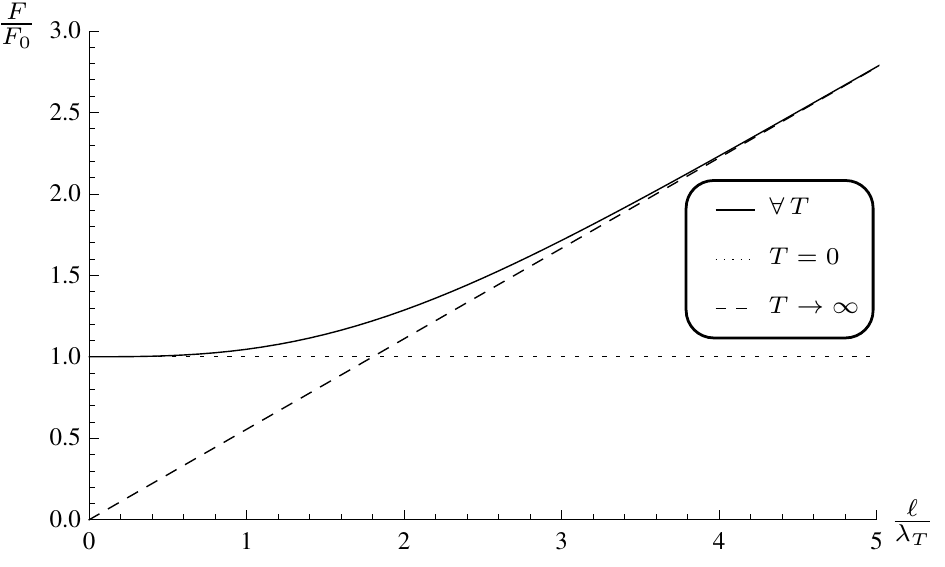}
\caption{The left and right figures are the large and short distance limit of the Casimir force between perfect metal spheres as a function of $\frac{d}{\lambda_{T}}$ (left) and of $\frac{\ell}{\lambda_{T}}$ (right) compared with the energy at zero temperature respectively. The dotted curve is the quantum limit, the dashed curve is the classical limit, and the solid curve is the asymptotic finite-temperature Casimir energy. The nonmonotonic behavior of the force with temperature in the large distance limit results from the negative slope of the solid curve at constant separation and dissappears in the short distance case.}
\label{Fuerza_2_esferas_metalicas}
\end{center}
\end{figure}

In this section we calculate the Casimir force between spheres. In the large distance limit, the asymptotic Casimir force $F = -\partial_{d}E$ can be written in terms of the adimensional Casimir energy as
\begin{equation}\label{Fuerza_adimensional}
F_{ad}(z) = \frac{2\pi}{\hbar c}\frac{d^{8}}{R^{6}}F = 7E_{ad}(z) - z\partial_{z}E_{ad}(z),
\end{equation}
where $z = \frac{d}{\lambda_{T}}$. In Fig. \ref{Fuerza_2_esferas_metalicas}(left), the adimensional asymptotic force between the perfect metal spheres compared with the zero-temperature force is plotted as a function of $\frac{d}{\lambda_{T}}$ for the large and short distance limits. Here, for constant temperature, we observe nonmonotonic behavior of the force with the adimensional parameter $\frac{d}{\lambda_{T}}$ in the large distance limit, which dissappears in the short distance case.


It is easy to verify that the force behaves monotonically with separation, and the nonmonotonicity of the entropy with separation implies nonmonotonic behavior of the force with temperature (but not with the distance), because
\begin{equation}\label{Relacion_Fuerza_Entropia}
\frac{\partial F}{\partial T} = - \frac{\partial^{2} E}{\partial T\partial d} = \frac{\partial S}{\partial d},
\end{equation}
so the appearance of negative slopes of the entropy with separation implies nonmonotonicity of the Casimir force with temperature, despite the attractive force for all separations and temperatures.

As observed in Fig. \ref{Fuerza_2_esferas_metalicas}, the force between the spheres is always attractive, but it is not always monotonic with temperature; asymptotically, for any given temperature, there exists a range of separations for which the force decreases with temperature. Then we have obtained a nonmonotonic behavior of the Casimir force with temperature between compact objects. Nonmonotonicities of the force between a plate and cylinder and between a plate and sphere were already obtained in Ref.~\refcite{Metodo_Caminos_Esfera_Placa_Escalar}, but in that case the nonmonotonicity already appears for the scalar field. Nonmonotonicity does not appear between spheres for the scalar field; this is a characteristic effect of the electromagnetic field, because cross-polarization terms of the Casimir energy are essential for this nonmonotonicity to appear.

In the Short distance limit, the force is obtained from the PFA energy given in Eq.~\eqref{PFA_8} as
\begin{equation}\label{PFA_14}
F^{PFA}_{T} = - \frac{k_{B}TR}{4\ell^{2}}\sum_{m=1}^{\infty}{\sum_{n=0}^{\infty}}'\frac{e^{- 2nm\ell\lambda_{T}^{-1}}}{m^{3}}\left(1 + 2nm\ell\lambda_{T}^{-1}\right).
\end{equation}
This result is shown in Fig. \ref{Fuerza_2_esferas_metalicas}(right), where er observe that the Casimir force is monotonous with the temperature in the short distance case.

\section{Thermodynamical Consequences}\label{Sect:Thermodynamic}
In this section we discuss the thermodynamical consequences of the obtained results. In this article we have obtained the large and short separation limit of the Casimir energy and entropy for two perfect metal spheres. At any nonzero fixed temperature, we observe an interval of separations for which the entropy is negative, at zero temperature the entropy is always zero, while it is not the minimum of the entropy.

In addition, the Casimir force is attractive for all separations and temperatures. So, we could naively think that we have possible violations of the second and third laws of thermodynamics, due to the existence of processes where the entropy of the system tends to increase and to the negative entropy intervals at finite temperature and distances, respectively.

According to the Krein formula\cite{PRB_Negative_Entropy_2_Spheres,Wirzba}, we know that the Helmholtz free energy of the electromagnetic field has three independent additive contributions: one is from the thermal bath, being proportional to the volume of the space\cite{Spectra_Finite_Systems}. Another is the sum of contributions of objects immersed in the bath considered as isolated objects, each contribution being also a function of the volume and surface of each object\cite{Spectra_Finite_Systems}. The third is of geometrical nature, which we could call the Casimir part of the Helmholtz free energy and that we actually calculate in Eq. \eqref{Energy_T_finite}. 

Considering the whole system, the nonmonotonicity of the Casimir entropy with temperature is compensated by the contribution of the vacuum, because one scales with the global volume\cite{Spectra_Finite_Systems} and the other with the separation between the spheres, so the behavior of the global entropy is dominated by the thermal bath contribution and the minimum is reached in the zero temperature limit, therefore there is not a violation of the third law. 


In addition, the second law of thermodynamics states that global entropy must increase for any process, but only in closed systems. As we are working in the canonical ensemble, we are implicitly assuming that there exists an external reservoir which keeps our system at constant temperature, so the system is not isolated and its entropy can increase or decrease without violation of the second law. In the canonical ensemble, the condition equivalent to the second law is that the global Helmholtz free energy must decrease for any process, and this is true for the studied system. Therefore, the appearance of nonmonotonic entropy behavior in the canonical ensemble just implies nonmonotonic behavior of the force with temperature, as seen in Eq. \eqref{Relacion_Fuerza_Entropia}.

\section{Conclusions}
In this article we have obtained the large and short separation limit of the Casimir energy and entropy for two perfect metal spheres. At any nonzero fixed temperature, we observe an interval of separations for which entropy is negative, while at zero temperature the entropy is always zero. We showed numerically that there exists a minimum separation between the spheres for which the negative entropy interval disappear. In Sect. \ref{Sect:Thermodynamic}, having into account the complete thermodynamical system, that the system is described by the canonical ensemble, and with the help of the Krein formula\cite{PRB_Negative_Entropy_2_Spheres,Wirzba} and Weyl formula\cite{Spectra_Finite_Systems}, we have demonstrated that there are not violations of second and third laws respectively. Therefore, the appearance of nonmonotonic entropy behavior in the canonical ensemble just implies nonmonotonic behavior of the Casimir force with temperature, as seen in Eq. \eqref{Relacion_Fuerza_Entropia}, which is attractive for all separations and temperatures. While this proceeding was on review process, a related work was published\cite{Teo_PFA_spheres}.

\section*{Acknowledgments}

The author acknowledge helpful discussions with R.~Brito, J.~Parrondo, J.~Horowitz, E.~Rold\'an, L.~Dinis and T.~Emig. This research is supported by Projects MOSAICO, UCM (Grant No. PR34/07-15859),
MODELICO (Comunidad de Madrid), ENFASIS (Grant No. FIS2011-22644, Spanish Government), and a FPU MEC grant.



\begin{thebibliography}{00}  
\bibitem{Casimir Placas Paralelas} H.B.G. Casimir, \textit{Proc. K. Ned. Akad. Wet.} \textbf{51}, 793 (1948).
\bibitem{EGJK}T. Emig, N. Graham, R.L. Jaffe, and M. Kardar, \textit{Phys. Rev. Lett.} \textbf{99}, 170403 (2007). 
\bibitem{Review-Jamal-Emig}S.J. Rahi, T. Emig, N. Graham, R.L. Jaffe, and M. Kardar, \textit{Phys. Rev. D} \textbf{80}, 085021 (2009). 
\bibitem{Multiscattering_Lambrecht}A. Lambrecht P. A. Maia Neto and S. Reynaud. \textit{New Journal of Physics} \textbf{8}, 243 (2006). 
\bibitem{Multiscattering_Milton}K. A. Milton and J. Wagner, \textit{J. Phys. A: Math. Theor.} \textbf{41}, 155402 (2008). 
\bibitem{Metodo_Caminos_Esfera_Placa_Escalar}A. Weber and H. Gies, \textit{Phys. Rev. Lett.} \textbf{105}, 040403 (2010). 
\bibitem{Entropias_negativas_placas_Drude}V.B. Bezerra, G.L. Klimchitskaya, and V.M. Mostepanenko, \textit{Phys. Rev. A} \textbf{65}, 052113 (2002). 
\bibitem{Bordag_Entropia_Esfera_Placa}M. Bordag and I. G. Pirozhenko. \textit{Phys. Rev. D} \textbf{82}, 125016 (2010). 
\bibitem{Canaguier-Durand_Caso_Esfera_Placa_PRA}A. Canaguier-Durand, P.A. Maia Neto, A. Lambrecht, and S. Reynaud, \textit{Phys. Rev. A} \textbf{82}, 012511 (2010). 
\bibitem{PRB_Negative_Entropy_2_Spheres}P. Rodriguez-Lopez, \textit{Phys. Rev. B} \textbf{84}, 075431 (2011). 
\bibitem{Brevik}S.A. Ellingsen, I. Brevik, J.S. H\o{}ye and K.A. Milton, \textit{Phys. Rev. E} \textbf{78}, 021117 (2008). 
\bibitem{Wittmann}R.C. Wittmann, \textit{IEEE Trans. Antennas Propag.} \textbf{36}, 8 (1988).
\bibitem{Teo_PFA}L.P. Teo, \textit{Phys. Rev. D} \textbf{84}, 025022 (2011). 
\bibitem{Post_PFA_original}C. D. Fosco, F. C. Lombardo and F. D. Mazzitelli, \textit{Phys. Rev. D} \textbf{84}, 105031 (2011).
\bibitem{Post_PFA_Emig}G. Bimonte, T. Emig, R. L. Jaffe and M. Kardar, \textit{EPL} \textbf{97}, 50001 (2012).
\bibitem{Ingold}G.-L. Ingold, A. Lambrecht, and S. Reynaud, \textit{Phys. Rev. E} \textbf{80}, 041113 (2009). 
\bibitem{Zandi_Emig_Placa_Esfera_varios_modelos}R. Zandi, T. Emig, and U. Mohideen. \textit{Phys. Rev. B} \textbf{81}, 195423 (2010). 
\bibitem{Wirzba}A. Wirzba, \textit{J. Phys. A: Math. Theor.} \textbf{41} 164003 (2008). 
\bibitem{Spectra_Finite_Systems} H.P. Baltes and E.R. Hilf, \textit{Spectra of Finite Systems}, (Birkh\"{a}user Boston, 1980).
\bibitem{Teo_PFA_spheres}L.P. Teo, \textit{Phys. Rev. D} \textbf{85}, 045027 (2012). 






\end{thebibliography}
\end{document}